\documentstyle [12pt]{article}
\def\be{\begin{equation}}
\def\ee{\end{equation}}

\begin{document}

\title{$p$--Adic pseudodifferential operators \\ and $p$--adic wavelets}
\author{S.V.Kozyrev}
\maketitle

\begin{abstract}
We introduce a new wide class of $p$--adic pseudodifferential
operators. We show that the basis of $p$--adic wavelets is the
basis of eigenvectors for the introduced operators.
\end{abstract}

\section{Introduction}

In the present paper we introduce the family of pseudodifferential
operators, which generate $p$--adic diffusions. The constructed
operators, unlike the Vladimirov operator of $p$--adic fractional
derivation \cite{VVZ}, will not be diagonalizable by the Fourier
transform. These operators will be diagonal in the basis of
$p$--adic wavelets, introduced in \cite{wavelets}. Therefore we
will call these operators non translationally invariant (i.e., non
commutative with the operators of shift).

In the present paper we consider only the problem of existence of
the mentioned operators. To show that the introduced operators $T$
generate diffusions, one has to show that the corresponding
evolutions $e^{-tT}$ exist, conserve the cone of positive
functions and the total number of particles. Formally these
conditions are satisfied, the rigorous proof will be considered
elsewhere.

Ultrametric diffusions were considered in \cite{VVZ}, \cite{AlKa},
\cite{Kochubei1}, \cite{Kochubei2}, \cite{Kochubei3}.
In \cite{ABK}, \cite{ABKO}, \cite{ABO} ultrametric diffusions
were applied for description of the relaxation processes
in disordered systems, for instance, in spin glasses and proteins.
In the present paper we
describe diffusions with more general generators, compared to the
considered in the mentioned above literature.

The field of $p$--adic numbers is the simplest example of an
ultrametric space. $p$--Adic mathematical physics, see
\cite{VVZ}--\cite{Carlucci1}, studies the models of
mathematical physics with the help of $p$--adic analysis.

The results of the present paper have the following physical applications.

In the works \cite{ABK}, \cite{PaSu} $p$--adic analysis was
applied to describe the replica symmetry breaking, for discussion
of the replica approach see \cite{MPV}. It was shown that the
Parisi matrix describing the replica symmetry breaking (before the
$n\to 0$ limit) is a discrete analogue of a $p$--adic
pseudofifferential operator. In the present paper we introduce a
new family of $p$--adic pseudodifferential operators. The discrete
analogues of the introduced operators will be the new examples of
the replica matrices. The application of these replica matrices to
disordered models (for which the $n\to 0$ limit for the introduced
operators have to be developed) would be an interesting problem.

Another possible application of the approach of the present paper is
the application to random walk on hierarchical landscapes of energy,
relaxation phenomena in complex systems and basin to basin kinetics, see
\cite{Frauenfelder2} -- \cite{Frauenfelder3}, \cite{Becker1}.

The structure of the present paper is as follows.

In Section 2 we remind some results from $p$--adic analysis and
describe the basis of $p$--adic wavelets.

In Section 3 we introduce the family of pseudodifferential
operators and prove that these operators are diagonal in the basis
of $p$--adic wavelets.

In Section 4 we discuss in more details some particular examples
of pseudodifferential operators and compute the asymptotics of
population of unit disk for the corresponding ultrametric
diffusion.

In Section 5 we in more details discuss the possible physical applications
of the results of the present paper.

\section{$p$--Adic analysis}

Remind some notations used in $p$--adic analysis, see \cite{VVZ}
for details.

An ultrametric space is a metric space with the metrics
$\|\cdot\|$ satisfying the strong triangle inequality
$$
\|xy\| \le \max (\|xz\|,\|yz\|),\qquad \forall z
$$

The simplest example of ultrametric space is the field $Q_p$ of
$p$--adic numbers. The field $Q_p$ of $p$--adic numbers is the
completion of the field of rational numbers  $Q$ with respect to
$p$--adic norm on $Q$. This norm is defined in the following way.
An arbitrary rational number $x$ can be written in the form
$x=p^{\gamma}\frac{m}{n}$ with $m$ and $n$ not divisible by $p$.
The $p$--adic norm of the rational number
$x=p^{\gamma}\frac{m}{n}$ is equal to $|x|_p=p^{-\gamma}$. The
strong triangle inequality for $p$--adic numbers takes the form
$$
|x+y|_p \le \max (|x|_p,|y|_p).
$$

$p$--Adic numbers are in one to one correspondence with the
(convergent in $p$--adic norm) series of the form \be\label{x}
x=\sum_{\gamma=\gamma_0}^{\infty}x_i p^i,\qquad x_i=0,\dots,p-1
\ee The ring $Z_p$ of $p$--adic integers contains $p$--adic
numbers of the form
$$
x=\sum_{\gamma=0}^{\infty}x_i p^i,\qquad x_i=0,\dots,p-1
$$
and the group $Q_p/Z_p$ we identify with the set containing
fractions
$$
x=\sum_{\gamma=\gamma_0}^{-1}x_i p^i,\qquad x_i=0,\dots,p-1
$$

The  Vladimirov operator of $p$--adic fractional derivation is
defined as follows \cite{VVZ} \be\label{Dalpha} D^{\alpha}
f(x)=\frac{1}{\Gamma_p(-\alpha)}
\int_{Q_p}\frac{f(x)-f(y)}{|x-y|_p^{1+\alpha}}dy \ee where $dy$ is
the Haar measure on $Q_p$ and $\Gamma_p(-\alpha)$ is a constant.
This operator is diagonalizable by the $p$--adic Fourier transform
$F$:
$$
D^{\alpha} f(x)=F^{-1} \circ |k|_p^{\alpha}\circ F f(x)
$$

$p$--Adic wavelets were introduced in \cite{wavelets}, where
the following theorem was proven.

\bigskip

\noindent {\bf Theorem}.\quad {\sl The set of functions $\{\psi_{\gamma j n}\}$:
\begin{equation}\label{basis}
\psi_{\gamma j n}(x)=
p^{-{\gamma\over 2}} \chi(p^{\gamma-1}j x) \Omega(|p^{\gamma} x-n|_p),\quad
\gamma\in {\bf Z},\quad n\in Q_p/Z_p,\quad j=1,\dots,p-1
\end{equation}
is an orthonormal basis in $L^2(Q_p)$ of eigenvectors of the operator $D^{\alpha}$:
\begin{equation}\label{eigenvalues}
D^{\alpha}\psi_{\gamma j n}= p^{\alpha(1-\gamma)}\psi_{\gamma j n}
\end{equation}

}

\bigskip

Here $\Omega(x)$ is the indicator of the interval $[0,1]$, which
implies for the characteristic function of $p$--adic disk the
expression
$$
\Omega(|x|_p)=\left\{
\begin{array}{rc}
1,&|x|_p\le 1\;,\\
0,&|x|_p>1\;.\\
\end{array}
\right.
$$
The function $\chi(x)$ is the complex valued character of the
$p$--adic argument which for $x$ given by (\ref{x}) is defined as
$$
\chi(x)=\exp \left(2\pi i \sum_{\gamma=\gamma_0}^{-1}x_i p^i\right)
$$

The theorem above shows that the $p$--adic wavelets are the basis
of eigenvectors for the operator (\ref{Dalpha}). The operator
(\ref{Dalpha}), see \cite{VVZ}, was considered as the generator of
the ultrametric diffusion. In the present paper we introduce a
more general class of $p$--adic pseudodifferential operators.
These operators can not be diagonalized by the $p$--adic Fourier
transform, but we will show, that the introduced operators are
still diagonalizable by the $p$--adic wavelet transform (i.e. the
$p$--adic wavelets are eigenvectors of the introduced operators).

$p$--Adic numbers are related to the Bruhat--Tits trees, see
\cite{Serre}. The Bruhat--Tits tree is the infinite tree (with the
infinite number of edges and vertices) where each vertex is
connected with exactly $p+1$ over vertices. $p$--Adic projective
line, which contains the field of $p$--adic numbers plus one
infinite point, is equivalent to the space of classes of
equivalence of semiinfinite paths in the Bruhat--Tits tree. Two
(semiinfinite) paths are equivalent if they coincide,
starting from some vertex.
For instance, the set of semiinfinite paths going
through the fixed vertex and through $p$ edges incident to the
vertex is homeomorphic to a $p$--adic disk.

This construction is the analogue of the Poincare model of the
Lobachevskii plane, where the Bruhat--Tits tree is the analogue of
the Lobachevskii plane, the $p$--adic (projective) line is the
analogue of the absolute, and the (unique) path in the
Bruhat--Tits tree connecting $x$ and $y$ in $Q_p$ is the analogue
of the geodesic line connecting the points at the absolute.

We introduce the level, or height of the vertex, in the following
way. We put in correspondence to every vertex in the Bruhat--Tits
tree the $p$--adic disk, corresponding to the following set of
semiinfinite paths in the tree which begin in the vertex. This set
contains all the paths going through the $p$ of $p+1$ edges
incident to the vertex (excluding one edge incident to the path
going through the infinite point of the absolute). We call the
level of the vertex the $\log_p$ of the radius of this disk.

For two points $x$, $y$ in $Q_p$ there exists the unique path in
the Bruhat--Tits tree connecting these two points. This path
contains the unique point $A(x,y)$ with the highest level.

\section{Construction of the pseudodifferential operators}

In \cite{OS}, \cite{Yoshino} examples of ultrametric diffusions on a
discrete set were considered.
In \cite{VVZ}, \cite{AlKa}, \cite{Kochubei1}--\cite{Kochubei3},
\cite{ABKO}--\cite{ABO}
examples of $p$--adic diffusions were investigated.

In the present section we introduce a family of $p$--adic
pseudodifferential operators, which generate ultrametric
diffusions on the field $Q_p$ of $p$--adic numbers. The existence
of corresponding diffusions (possibility to take the exponent of
the generator) will not be discussed in the present paper. The
introduced pseudodifferential operators will be the
generalizations of the Vladimirov operator.

Consider the operator
\be\label{generator}
T f(x)=\int T(x,y)(f(x)-f(y))dy \ee Here we take $x$, $y$ lying in
the field $Q_p$ of $p$--adic numbers.
We consider the operator $T$
as the operator in the space $L^2(Q_p)$ of complex valued square
integrable functions of $p$--adic argument.

In the following we will introduce some conditions for the kernel
$T(x,y)$. Operator of the form (\ref{generator}) satisfying these
conditions we call the generator of the ultrametric diffusion. We
show that under these conditions the operator (\ref{generator})
will be diagonal in the basis of $p$--adic wavelets, introduced in
\cite{wavelets}. This means that the assumption of ultrametricity
of the diffusion makes the problem of investigation of the
diffusion exactly solvable.

\bigskip

\noindent{\bf Definition 1}.\qquad {\sl We call the transition
probability (and the corresponding diffusion with the generator
given by (\ref{generator})) ultrametric, if
the transition probability $T(x,y)$
depends only on the highest point $A(x,y)$ lying at the path
between $x$ and $y$ in the Bruhat--Tits tree. }

\bigskip

Notice that the transition probability
$T(x,y)$ is symmetric with respect to $x$, $y$ and satisfies the
following property:

\bigskip

\noindent {\bf Lemma 2}.\qquad{\sl The function $T(x,y)$
is symmetric and positive with
respect to $y$, and, for fixed $x$, is locally constant
with respect to $y$ outside
any vicinity of $x$. Moreover, for fixed $x$, the following
condition is satisfied:
\be\label{lemma1} T(x,y)=\hbox{ const },
\qquad \hbox{ if  }\quad |x-y|_p=\hbox{ const } \ee Vice versa,
when the condition (\ref{lemma1}) is satisfied,
the function $T(x,y)$ satisfies the conditions of Definition 1.}

\bigskip

It is easy to see that the Vladimirov operator and similar
operators with translationally invariant kernels $T(x,y)$ (which
depend only on the diference $x-y$ and therefore, under the
corresponding conditions of integrability of the kernel, can be
diagonalized by the Fourier transform), satisfy the conditions of
lemma 2. In the present paper we introduce the examples of the
generators with non translationally invariant kernels.

The following theorem gives the general form of the kernel for the
operator (\ref{generator}).

\bigskip

\noindent {\bf Theorem 3}.\qquad {\sl The function of the form
\be\label{lemma3} T(x,y)= \sum_{\gamma n} T^{(\gamma n)}
\delta_{1,|p^{\gamma}x-p^{\gamma}y|_p} \Omega(|p^{\gamma}x-n|_p)
\ee where $T^{(\gamma n)}\ge 0$, $\gamma\in Z$, $n\in Q_p/Z_p$
satisfies the conditions of lemma 2.

Moreover, an arbitrary function satisfying (\ref{lemma1})
can be represented in the form (\ref{lemma3}).
}

\bigskip

\noindent{\it Proof}\qquad
Positivity of $T(x,y)$ is obvious.

Let us prove the symmetricity of $T(x,y)$. We have
\be\label{T-T}
T(x,y)-T(y,x)=
\sum_{\gamma n} T^{(\gamma n)} \delta_{1,|p^{\gamma}x-p^{\gamma}y|_p}
\left(\Omega(|p^{\gamma}x-n|_p)-\Omega(|p^{\gamma}y-n|_p)\right)
\ee
Let us prove that this expression is equal identically to zero.
Consider the case when $x$ is such that the following characteristic function
is non--zero:
$$
\Omega(|p^{\gamma}x-n|_p)=1
$$
This implies
\be\label{x-n}
|x-p^{-\gamma}n|_p\le p^{\gamma}
\ee
If
$$
\delta_{1,|p^{\gamma}x-p^{\gamma}y|_p}\ne 0
$$
then
\be\label{x-y}
|x-y|_p= p^{\gamma}
\ee
Formulas (\ref{x-n}), (\ref{x-y}) and ultrametricity imply that
$$
\Omega(|p^{\gamma}y-n|_p)=1
$$
Therefore the corresponding terms in (\ref{T-T}) cancel.
This proves that $T(x,y)$ given by (\ref{lemma3}) is symmetric.

Prove that $T(x,y)$ given by (\ref{lemma3}) satisfies (\ref{lemma1}).
Fix $x\in Q_p$. Then for $y$ lying at the sphere with the center in $x$
and the radius $p^{\gamma}$ we have
$$
\delta_{1,|p^{\gamma}x-p^{\gamma}y|_p}=1
$$
and
$$
\Omega(|p^{\gamma}x-n|_p)
$$
will be equal to 0 or 1 depending on whether $p^{-\gamma}n$ lies in the disk
$$
\{y: |x-y|_p\le p^{\gamma}\}
$$
or no.
Therefore $T(x,y)$ will be a constant on the considered sphere
and (\ref{lemma1}) will be satisfied.

This proves that $T(x,y)$ satisfying the conditions of the present
lemma will satisfy lemma 2.

Vice versa, it is easy to see that the kernel (\ref{lemma3})
for $x$, $y$ lying in the disk with the center in $p^{-\gamma}n$
and the radius
$p^{\gamma}$, and satisfying $|x-y|_p=p^{\gamma}$, takes the
value $T^{(\gamma n)}$.

Since all the space $x,y\in Q_p\times Q_p$ is the disjoint union
of such a subsets, therefore, taking an arbitrary positive
$T^{(\gamma n)}$ we are able to construct an arbitrary kernel
satisfying (\ref{lemma1}). This finishes the proof of the theorem.

\bigskip

In the next theorem we compute the eigenvalues of the generator of
ultrametric diffusion.

\bigskip

\noindent {\bf Theorem 4}.\qquad {\sl Let the kernel
(\ref{lemma1}) satisfy the condition of convergence of all the
integrals in (\ref{lambdagn}) for any $\gamma$, $n$. Then the
operator (\ref{generator}) is a well defined operator in
$L^2(Q_p)$ with the dense domain and the $p$--adic wavelets
$\psi_{\gamma j n}$ are eigenvectors for the ultrametric
diffusion: \be\label{lemma2} T\psi_{\gamma j n}=\lambda_{\gamma
n}\psi_{\gamma j n} \ee with the eigenvalues \be\label{lambdagn}
\lambda_{\gamma n}=
\int_{|n-p^{\gamma}y|_p>1}T({p^{-\gamma}n,y})dy+
p^{\gamma}T({p^{-\gamma}n,p^{-\gamma}(n+1)}) \ee

}

\bigskip

\noindent{\it Proof}\qquad To prove the present theorem we use
Lemma 2. Consider the wavelet $\psi_{\gamma j n}$. Then
$$
T\psi_{\gamma j n}(x)=\int T(x,y)\left(\psi_{\gamma j
n}(x)-\psi_{\gamma j n}(y) \right)dy
$$

Consider the following cases.

1) Let $|p^{\gamma}x-n|_p>1$. Then Lemma 2 implies
$$
T\psi_{\gamma j n}(x)=\int T(x,y)\psi_{\gamma j n}(x)dy-
T({x,n})\int \psi_{\gamma j n}(y)dy=0
$$

2) Let $|p^{\gamma}x-n|_p \le 1$. Then again by Lemma 2
$$
T\psi_{\gamma j n}(x)=
\left(\int_{|p^{\gamma}x-p^{\gamma}y|_p>1}+
\int_{|p^{\gamma}x-p^{\gamma}y|_p=1}+
\int_{|p^{\gamma}x-p^{\gamma}y|_p<1}\right)T(x,y)
(\psi_{\gamma j n}(x)-\psi_{\gamma j n}(y))dy=
$$
$$
=
\left(\int_{|p^{\gamma}x-p^{\gamma}y|_p>1}+
\int_{|p^{\gamma}x-p^{\gamma}y|_p=1}\right)T(x,y)
(\psi_{\gamma j n}(x)-\psi_{\gamma j n}(y))dy=
$$
$$
=
\psi_{\gamma j n}(x)\int_{|p^{\gamma}x-p^{\gamma}y|_p>1}T(x,y)dy+
\int_{|p^{\gamma}x-p^{\gamma}y|_p=1}T(x,y)
(\psi_{\gamma j n}(x)-\psi_{\gamma j n}(y))dy=
$$
$$
=
\psi_{\gamma j n}(x)\int_{|n-p^{\gamma}y|_p>1}T({p^{-\gamma}n,y})dy+
\sum_{l=0}^{p-1}\int_{|z|_p<p^{\gamma}}
T({x,x+p^{-\gamma}l+z})
(\psi_{\gamma j n}(x)-\psi_{\gamma j n}(x+p^{-\gamma}l+z))dz=
$$
$$
=
\psi_{\gamma j n}(x)\int_{|n-p^{\gamma}y|_p>1}T({p^{-\gamma}n,y})dy+
p^{\gamma-1}\sum_{l=1}^{p-1}
T({x,x+p^{-\gamma}l})
(\psi_{\gamma j n}(x)-\psi_{\gamma j n}(x+p^{-\gamma}l))
$$
The last equality follows from Lemma 2 and the local constance of
$\psi_{\gamma j n}$.

Using the identity
$$
\psi_{\gamma j n}(x+p^{-\gamma}l)=
p^{-{\gamma\over 2}} \chi(p^{\gamma-1}j (x+p^{-\gamma}l))
\Omega(|p^{\gamma} (x+p^{-\gamma}l)-n|_p)=
\chi(p^{-1}j l)\psi_{\gamma j n}(x)
$$
and Lemma 2, we obtain
$$
T\psi_{\gamma j n}(x)=\psi_{\gamma j n}(x)\left(
\int_{|n-p^{\gamma}y|_p>1}T({p^{-\gamma}n,y})dy+
p^{\gamma-1}T({p^{-\gamma}n,p^{-\gamma}(n+1)})
\sum_{l=1}^{p-1}
(1-\chi(p^{-1}j l))
\right)
$$
Since
$$
\sum_{l=1}^{p-1}
(1-\chi(p^{-1}j l))=p
$$
we obtain
$$
T\psi_{\gamma j n}(x)=\psi_{\gamma j n}(x)
\left(
\int_{|n-p^{\gamma}y|_p>1}T({p^{-\gamma}n,y})dy+
p^{\gamma}T({p^{-\gamma}n,p^{-\gamma}(n+1)})
\right)
$$
which gives (\ref{lambdagn}).

Therefore the operator $T$ is well defined on the basis in
$L^2(Q_p)$ that finishes the proof of the theorem.
Note that convergence of all the integrals used in the proof
is guaranteed by convergence of (\ref{lambdagn}).

\bigskip

The next corollary gives a simple representation for the
eigenvalues of the generator with the kernel
(\ref{lemma3}).

\bigskip

\noindent {\bf Corollary 5}.\qquad {\sl Let the following series
converge for any $n$: \be\label{seriesconverge}
\sum^{\infty}_{\gamma=0} p^{\gamma} T^{(\gamma 0)}<\infty
\ee Then the operator (\ref{generator}) corresponding to the
kernel (\ref{lemma3}) is well defined in $L^2(Q_p)$,
diagonal in the basis of $p$--adic wavelets and have the
eigenvalues
\be\label{lemma4} \lambda_{\gamma n}=
p^{\gamma}T^{(\gamma n)}+ (1-p^{-1})
\sum^{\infty}_{\gamma'=\gamma+1} p^{\gamma'}
T^{\left(\gamma',p^{\gamma'-\gamma} n\right)} \ee }

\bigskip

Note that condition of convergence of the series
(\ref{seriesconverge}) is equivalent to convergence of the
integral (\ref{lambdagn}).

\bigskip

\noindent{\it Proof}\qquad
Substituting (\ref{lemma3}) into (\ref{lambdagn}) we get
$$
\lambda_{\gamma n}=
\int_{|n-p^{\gamma}y|_p>1}
\sum_{\gamma' n'} T^{(\gamma' n')} \delta_{1,|p^{\gamma'}p^{-\gamma}n-
p^{\gamma'}y|_p}\Omega(|p^{\gamma'}p^{-\gamma}n-n'|_p)dy+
$$ $$
+p^{\gamma}
\sum_{\gamma' n'} T^{(\gamma' n')}
\delta_{1,|p^{\gamma'}p^{-\gamma}n-p^{\gamma'}p^{-\gamma}(n+1)|_p}
\Omega(|p^{\gamma'}p^{-\gamma}n-n'|_p)=
$$ $$
=
\sum_{\gamma' n'} T^{(\gamma' n')} p^{\gamma'}(1-p^{-1})\theta(\gamma'-\gamma)
\Omega(|p^{\gamma'}p^{-\gamma}n-n'|_p)+
p^{\gamma}
\sum_{n'} T^{(\gamma n')}
\Omega(|n-n'|_p)
=
$$ $$
=p^{\gamma}T^{(\gamma n)}+
(1-p^{-1}) \sum^{\infty}_{\gamma'=\gamma+1}
p^{\gamma'} T^{\left(\gamma',p^{\gamma'-\gamma} n\right)}
$$
Here we use the properties
$$
\int_{|n-p^{\gamma}y|_p>1}\delta_{1,|p^{\gamma'}p^{-\gamma}n-
p^{\gamma'}y|_p} dy=p^{\gamma'}(1-p^{-1})\theta(\gamma'-\gamma),\quad
\theta(\gamma)=0,\gamma\le 0,  \theta(\gamma)=1,\gamma>0
$$
$$
\delta_{1,|p^{\gamma'}p^{-\gamma}n-p^{\gamma'}p^{-\gamma}(n+1)|_p}
=\delta_{\gamma\gamma'}
$$
$$
\Omega(|m-n|_p)=\delta_{mn}, \qquad m,n \in Q_p/Z_p
$$
This finishes the proof of the corollary.

\bigskip

\noindent {\bf Remark}.\qquad  This corollary implies the
recurrent formulas
$$
\lambda_{\gamma-1,p^{-1} n}-\lambda_{\gamma n}=
p^{\gamma-1}T^{(\gamma-1, p^{-1}n)}-p^{\gamma-1}T^{(\gamma n)}
$$
\be\label{recurrent} T^{(\gamma n)}=T^{(\gamma-1,p^{-1} n)}+p^{1-\gamma}
(\lambda_{\gamma n}-\lambda_{\gamma-1,p^{-1} n}) \ee For instance the
sequence $\{\lambda_{\gamma n}\}$ for fixed $n$ is non--negative
and non--increasing with the increase of $\gamma$.

\section{Examples of the pseudodifferential operators and properties of the diffusions}

Calculate particular examples of the kernels of the kind (\ref{lemma3}).

\medskip

\noindent{\large Example 1}.\qquad
Let us consider the translationally invariant case. In this case
$$
T^{(\gamma n)}=T^{(\gamma)}
$$
and the summation over $n$ in (\ref{lemma3}) gives
$$
\sum_n  \Omega(|p^{\gamma}x-n|_p)=1
$$
Denoting
$$
T^{(\gamma )}=f(p^{\gamma})
$$
we obtain for the kernel
$$
T(x,y)=
\sum_{\gamma } T^{(\gamma )} \delta_{1,|p^{\gamma}x-p^{\gamma}y|_p}=
f(|x-y|_p)
$$
Therefore the choice
\be\label{degree}
T^{(\gamma )}=p^{-\gamma(1+\alpha)}
\ee
corresponds to the Vladimirov operator, when the kernel is given by
$$
T(x,y)={1\over |x-y|_p^{1+\alpha}}
$$

Formula (\ref{lemma4}) in the considered case (\ref{degree}) takes the form
\be\label{lambdadegree}
\lambda_{\gamma n}=
p^{\gamma}T^{(\gamma )}+
(1-p^{-1}) \sum^{\infty}_{\gamma'=\gamma+1}
p^{\gamma'} T^{(\gamma' )}=
$$
$$
=p^{\gamma}p^{-\gamma(1+\alpha)}+
(1-p^{-1}) \sum^{\infty}_{\gamma'=\gamma+1}
p^{\gamma'} p^{-\gamma'(1+\alpha)} =
p^{-\gamma\alpha}{1-p^{-\alpha-1}\over 1-p^{-\alpha}}
\ee

\medskip

\noindent{\large Example 2}.\qquad
Consider the case when
\be\label{ex2Tgn}
T^{(\gamma n)} =
f(p^{\gamma})g(|p^{-\gamma}n-n_0|_p)
\ee
This implies for the kernel
$$
T(x,y)=
\sum_{\gamma } f(p^{\gamma}) \delta_{1,|p^{\gamma}x-p^{\gamma}y|_p}
\sum_{n}g(|p^{-\gamma}n-n_0|_p)
\Omega(|p^{\gamma}x-n|_p)=
$$
$$
=f(|x-y|_p)\sum_{\gamma }  \delta_{p^{\gamma},|x-y|_p}
\sum_{n}g(|p^{-\gamma}n-n_0|_p)
\Omega(|p^{\gamma}x-n|_p)
$$

One can check that the series
$$
\sum_{n\in Q_p/Z_p}g\left({|n-n_0|_p}\right)
\Omega\left({|x-n|_p}\right)
$$
equals to $g\left(|x-n_0|_p\right)$ for $|x-n_0|_p>1$ and to
$g(0)$ for $|x-n_0|_p\le 1$:
$$
\sum_{n\in Q_p/Z_p}g\left({|n-n_0|_p}\right)
\Omega\left({|x-n|_p}\right)=g\left(|x-n_0|_p\right)
(1-\Omega\left({|x-n_0|_p}\right))
+g(0)\Omega\left({|x-n_0|_p}\right)
$$
This implies
$$
\sum_{n\in Q_p/Z_p}g\left({|p^{-\gamma}n-n_0|_p}\right)
\Omega\left({|p^{\gamma}x-n|_p}\right)=$$ $$=
g\left(|x-n_0|_p\right)
(1-\Omega\left({|p^{\gamma}x-n_0|_p}\right))
+g(0)\Omega\left({|p^{\gamma}x-n_0|_p}\right)
$$

For the kernel this will give \be\label{ex2} T(x,y)=f(|x-y|_p)
\left( g\left(|x-n_0|_p\right)
(1-\Omega\left({||x-y|_px-n_0|_p}\right))
+g(0)\Omega\left({||x-y|_px-n_0|_p}\right) \right) \ee

Corollary 5 and (\ref{ex2Tgn}) imply for the eigenvalues
corresponding to diffusion operator with the kernel (\ref{ex2})
the following:
$$
\lambda_{\gamma n}=
p^{\gamma}f(p^{\gamma})g(|p^{-\gamma}n-n_0|_p)+
(1-p^{-1}) g(|p^{-\gamma}n-n_0|_p)\sum^{\infty}_{\gamma'=\gamma+1}
p^{\gamma'} f(p^{\gamma'})
$$

\bigskip

In the present section we investigate the relaxation of the indicator
function of the unit disk:
$$
\phi=\Omega(|x|_p)
$$
for the ultrametric diffusion with the kernel (\ref{lemma3}).

The indicator function has the following decomposition
over $p$--adic wavelets, see \cite{wavelets}
$$
\langle\Omega(|x|_p),\psi_{\gamma j n}\rangle= p^{-{\gamma\over
2}}\theta(\gamma)\delta_{n0},\qquad \theta(\gamma)=0,\gamma\le
0,\quad \theta(\gamma)=1,\gamma\ge 1
$$
$$
\Omega(|x|_p)=\sum_{j=1}^{p-1}\sum_{\gamma=1}^{\infty} p^{-{\gamma\over 2}}
\psi_{\gamma j 0}(x)
$$

Therefore the operator $T$ acts on $\Omega(|x|_p)$ as follows
$$
T\Omega(|x|_p)=
\sum_{j=1}^{p-1}\sum_{\gamma=1}^{\infty} p^{-{\gamma\over 2}}
\lambda_{\gamma0}\psi_{\gamma j 0}
$$
By (\ref{lemma4}) for the survival probability of the unit disk
with the center in zero we obtain \be\label{survival} S(t)=\langle
\Omega(|x|_p),e^{-tT}\Omega(|x|_p)\rangle=
(p-1)\sum_{\gamma=1}^{\infty} p^{-{\gamma}}
e^{-t\lambda_{\gamma0}}=
$$
$$
=(p-1)\sum_{\gamma=1}^{\infty} p^{-{\gamma}}
\exp(-t\left(
p^{\gamma}T^{(\gamma 0)}+
(1-p^{-1}) \sum^{\infty}_{\gamma'=\gamma+1}
p^{\gamma'} T^{(\gamma' 0)}
\right))
\ee

Analogously
$$
\langle\Omega(|p^{\gamma}x-n|_p),\psi_{\gamma' j n'}\rangle=
\langle\Omega(|p^{\gamma}x-n|_p),
p^{-{\gamma'\over 2}} \chi(p^{\gamma'-1}j x) \Omega(|p^{\gamma'} x-n'|_p)
\rangle=
$$
$$
=p^{-{\gamma'\over 2}}\theta(\gamma'-\gamma)\chi(p^{\gamma'-\gamma-1}j n)
p^{\gamma}\Omega(|p^{\gamma'-\gamma}n-n'|_p)
$$
which implies
\be\label{decompose}
\Omega(|p^{\gamma}x-n|_p)=p^{\gamma}\sum_{j=1}^{p-1}\sum_{\gamma'=\gamma+1}^{\infty}
p^{-{\gamma'\over 2}}\chi(p^{\gamma'-\gamma-1}j n)
\psi_{\gamma' j p^{\gamma'-\gamma}n}(x)
\ee
where we use the formula
$$
\Omega(|p^{\gamma'-\gamma}n-n'|_p)=\delta_{p^{\gamma'-\gamma}n,n'}
$$
Formula (\ref{decompose}) allows to compute the analogues of
(\ref{survival}) for displaced disks.

Note that $p^{\gamma'-\gamma}n=0$ in $Q_p/Z_p$ when $\gamma'$ is large enough.
Thus the series over the wavelets in (\ref{decompose}) contains
only the finite sum of the wavelets with the third index not equal
to zero. Therefore the asymptotic dependence on time of
$$
\langle \Omega(|p^{\gamma}x-n|_p),e^{-tT}\Omega(|p^{\gamma'}x-n'|_p)\rangle
$$
coincides with the asymptotics of (\ref{survival}).

\section{Conclusion}

In the present paper we construct the new family of $p$--adic
pseudodifferential operators $T$ and show that these operators are diagonal
in the basis of $p$--adic wavelets. This allows to compute analytically
the corresponding evolutions $e^{-tT}$.

The approach of the present paper may be applied to the problem
of investigation of the diffusion on hierarchical energy landscapes.
The concept of hierarchical energy landscape attracts a lot of
interest in connection with investigation of complex systems, in
particular, glasses, clusters, and proteins
\cite{Mezard1}-\cite{Becker1}.
Discuss this concept using the language of basis to basis kinetics
\cite{Becker1}. In the framework of this concept
a complex system is assumed to have a large
number of metastable configurations, corresponding to local minima of
energy. The local minima
are clustered in hierarchically nested basins of minima, namely,
each large basin consists of smaller basins, each of these consisting
of even smaller ones, and so on. In the other words,
the hierarchy of basins possesses ultrametric geometry. Finally, the basins
of minima are separated from one another by a hierarchically arranged set of
barriers, i.e.,  high barriers separate large basins, and smaller
basins within each larger one are separated by lower barriers.

The concept of hierarchical energy landscape may be considered as a
kind of approximation of a complex energy landscape by a
hierarchical energy landscape.

The dynamics of the system
with hierarchical energy landscape it is natural to describe by
the models of ultrametric diffusion.

It is commonly recognized that relevant
analytical tools should be developed for applying the concept
of  hierarchical energy landscape to the description of
relaxation phenomena in complex systems \cite{Frauenfelder2} --
\cite{Frauenfelder3}, \cite{Becker1}
(the concept of ``basin--to--basin kinetics'').

The present paper is devoted to the development of these analytical tools.
We show, that for some wide class of ultrametric diffusions the corresponding
generators ($p$--adic pseudodifferential operators)
may be diagonalized in the basis of $p$--adic wavlets,
introduced in \cite{wavelets}.
This means that the random walk on the corresponding
energy landscape may be investigated analytically.

\bigskip

\centerline{\bf Acknowledgements}

\smallskip

The author would like to thank
V.S.Vladimirov, V.A.Avetisov, I.V. Volovich, A.Yu.Khrennikov,
A.H.Bikulov and V.A.Osipov for discussions and valuable comments.
This work has been partly supported by INTAS YSF 2002--160 F2,
CRDF (grant UM1--2421--KV--02), The Russian Foundation for
Basic Research (projects 02--01--01084),
and by the grant for the support of scientific schools NSh 1542.2003.1.

\end{document}